\documentclass[letter]{jpsj3}
\usepackage{txfonts}

\title{Unusual Field-Insensitive Phase Transition and Kondo Behavior in SmTi$_2$Al$_{20}$}

\author{Ryuji Higashinaka\thanks{E-mail address: higashin@tmu.ac.jp}, Takuya Maruyama, Akihiro Nakama, Ryoichi Miyazaki, Yuji Aoki and Hideyuki Sato %\\
}
\inst{Graduate School of Science, Tokyo Metropolitan University, Hachioji, Tokyo 192-0397, Japan}
\abst{Magnetization, electrical resistivity and specific heat measurements were performed on high-quality single crystalline SmTi$_2$Al$_{20}$ (residual resistivity ratio $\sim$ 40) grown by Al self-flux method.
A Kondo-like $\log T$ dependence in the resistivity is observed below 50 K.
We discovered a field-insensitive phase transition at $T_{x}$ = 6.5 K and a field-insensitive heavy fermion behavior with the electronic specific heat coefficient $\gamma$ = 150 mJ/(K$^{2}$ mol).
Specific heat analysis reveals that the ground state is a $\Gamma_{8}$ quartet state and the Sm magnetic dipole moment $m_{{\rm Sm}}$ ($\sim 0.5 \mu_{{\rm B}}$ at $T \simeq$ 0) orders below $T_{x}$ in spite of the field-insensitive behavior.
Possible reasons for the field insensitiveness will be discussed.
}

\kword{caged structure, heavy fermion, magnetization, transport, specific heat}

\begin{document}
\maketitle
The strongly correlated electron systems with caged structure (e.g. filled skutterudite $RT_{4}X_{12}$) have attracted much attention because of the wide variety of interesting properties such as heavy fermion (HF) behavior, multipolar ordering, unconventional superconductivity (SC) and rattling motion of guest ions.
In particular Pr- and Sm-based series have been intensively investigated due to unconventional physical properties ascribed to multipolar degrees of freedom in their low energy excitations, such as the unconventional HFSC in PrOs$_{4}$Sb$_{12}$\cite{02Bauer}, a possible multipolar ordering in SmRu$_{4}$P$_{12}$\cite{02Matsuhira,05Yoshizawa,07Aoki} and a field-insensitive HF state in SmOs$_{4}$Sb$_{12}$\cite{05Sanada}.
\\
%\section{Introduction}
\quad Among the cage structure compounds, $RT_2X_{20}$ ($R$ = rare earth, $T$ = transition metal, $X$ = Zn, Al) compounds crystallizing in the cubic CeCr$_2$Al$_{20}$-type structure with the space group $Fd\bar{3}m$ (\#227) have attracted much attention recently\cite{07Torika,08Saiga,09Jia,09Yoshiuchi,11Onimaru}.
In these compounds, $R$ ion is surrounded by neighboring 16 $X$ atoms and the symmetry at $R$ site is a cubic point symmetry $T_{d}$\cite{68Krypy,95Niemann,97Nasch,98Thiede}.
Recent studies on Al-based compounds clarified that Pr$T_{2}$Al$_{20}$ ($T$ = Ti and V) show a HF behavior and a multipolar ordering attributed to the nonmagnetic $\Gamma_{3}$ ground state doublet \cite{10Sakai}.
On the other hand, there are little reports on Sm-based Sm$T_{2}$Al$_{20}$ although they are expected to show interesting properties observed in Sm-based filled skutterudite compounds as noted above.
In this work, we have succeeded in growing single crystals of SmTi$_{2}$Al$_{20}$ and have discovered unconventional physical properties. \\
%
%
%\section{Experimental}
\quad Single crystals of SmTi$_{2}$Al$_{20}$ have been prepared by the Al self-flux method from the starting elements, chips of Sm(3N), powders of Ti(5N) and grains of Al(4N) at the atomic ratio of 1:2:90.
These mixtures were placed in an alumina crucible and sealed in an evacuated quartz tube.
These sealed tubes were heated up to 1050 $^{\circ}$C, kept for 2 hours, then cooled down to 750 $^{\circ}$C for 45 hours.
The excess Al was spun off in a centrifuge.
Small amount of residual Al-flux on the crystal sufaces were removed by etching with 0.3vol\% NaOH in H$_{2}$O for a few hours in an ultrasonic bath.
Quality of the samples was checked using a powder x-ray diffractometer with Cu-K$\alpha$ radiation.
The lattice parameter of SmTi$_{2}$Al$_{20}$ was $a$ = 14.698(1) \AA.
This is in agreement with the previous report\cite{95Niemann}. 
DC magnetization measurements were carried out in a Quantum Design (QD) MPMS down to 2 K and up to 7 T.
Specific heat measurements were performed using a quasi-adiabatic method with a QD PPMS down to 2 K and up to 9 T and a dilution refrigerator down to 0.15 K and up to 8 T.
For these measurements, twenty pieces of the single crystals with a random crystalline-orientation were used.
Electrical resistivity and Hall coefficient were measured simultaneously using a standard AC five-probe technique with a QD PPMS down to 2 K and up to 9 T.
For these measurements, a plate like sample with (111) plane and with dimensions of 0.3 $\times$ 0.8 $\times$ 0.067 mm$^{3}$ was used.
\\
%
%
%\section{Results and Discussion}
%
\begin{figure}
\begin{center}
\includegraphics[width=\linewidth]{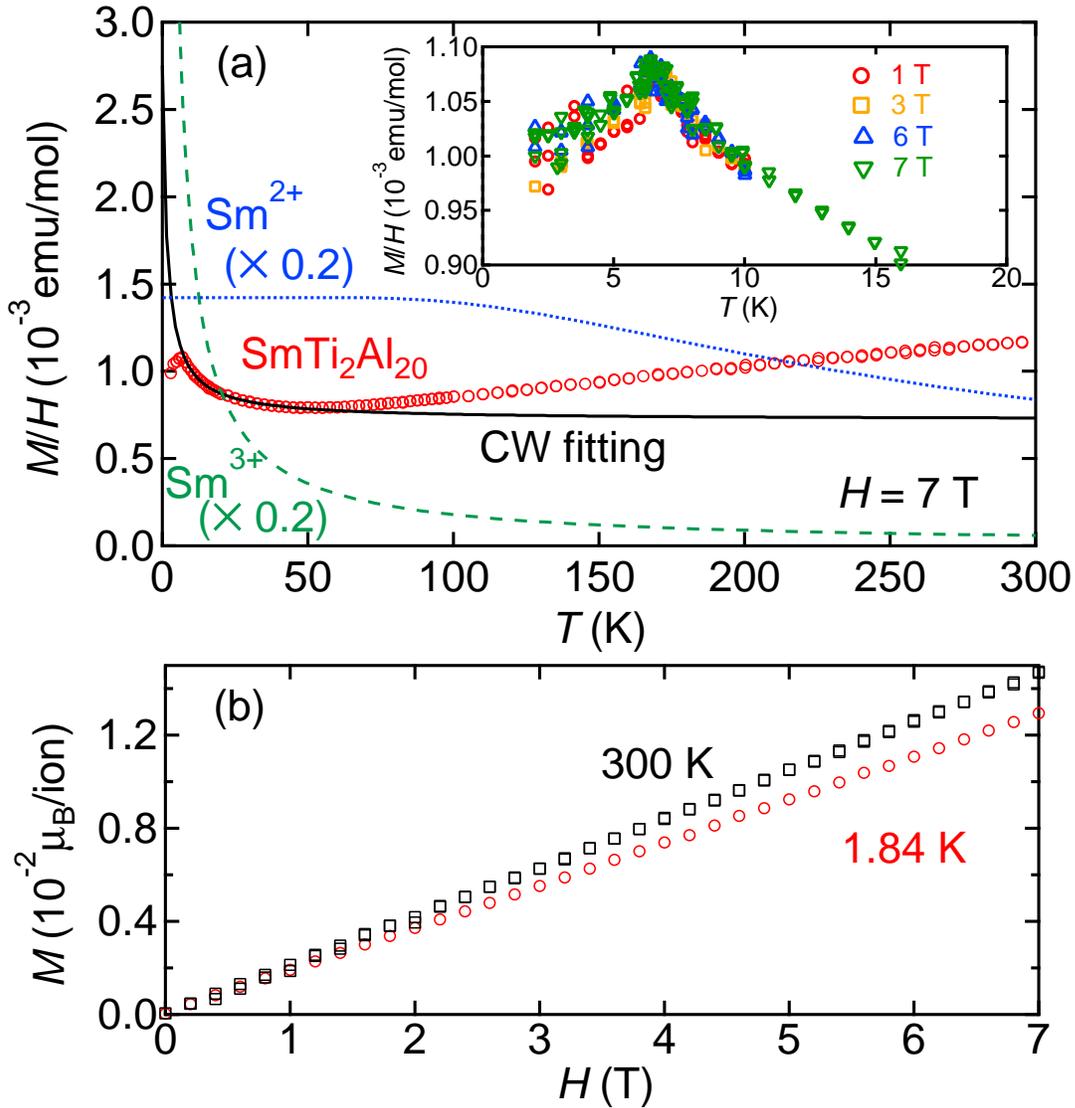}
\end{center}
\caption{(Color online) (a) $T$ dependence of DC susceptibility, $M/H$, for single crystalline SmTi$_{2}$Al$_{20}$ at 7 T (open circle) with random crystalline-orientation and a theoretical calculation divided by 5 for Sm$^{2+}$ (dotted line) and Sm$^{3+}$ ions (dashed line).
Solid line represents the CW fitting curve between 50 K and $T_{x}$. (see text for details)
Inset: low temperature part of $M/H$ at various fields.
(b) $H$ dependence of magnetization for SmTi$_{2}$Al$_{20}$ at 1.84 and 300 K.}
\label{f1}
\end{figure}
\quad In Fig. 1(a), we show the $T$ dependence of the DC susceptibility, $\chi$ (= $M/H$), measured at 7 T.
Compared with theoretically expected $\chi$ for Sm$^{2+}$ and Sm$^{3+}$ions, the observed behavior is different from both of the $T$ dependences.
Below room temperature (RT), $\chi$ decreases with decreasing temperature and shows a minimum at around 50 K (= $T_{{\rm min}}$).
Similar minimum structure in $\chi$ has been reported for SmAl$_{2}$\cite{73Wijn} and this could be attributed to the Van Vleck paramagnetic term derived from the mixing of the excited $J$ = 7/2 multiplet combining with crystalline electric field (CEF) effects and exchange interactions among Sm ions.
Below 50 K, it shows a Curie-like upturn followed by a sharp decrease below 6.5 K (= $T_{x}$).
In an enlarged view around $T_{x}$ in various magnetic fields [the inset of Fig.1 (a)], $T_{x}$ shows little field dependence.
The magnetization at room temperature and 1.84 K increase linearly up to 7 T [Fig1. (b)].
From a Curie Weiss (CW) fitting [$\chi = \chi_{0} + C/(T - \theta_{{\rm CW}})$] between $T_{x}$ and 50 K, $\chi_{0} = 7.2 \times 10^{-4}$ emu/mol, $C$ = 3.39 emu/mol K and $\theta_{{\rm CW}}$ = 1.7 K are obtained.
The estimated effective moment is 0.16 $\mu_{{\rm B}}$/Sm, which is only 20 \% of the free Sm$^{3+}$ ion value (0.84 $\mu_{{\rm B}}$/Sm).
\\
\begin{figure}
\begin{center}
\includegraphics[width=\linewidth]{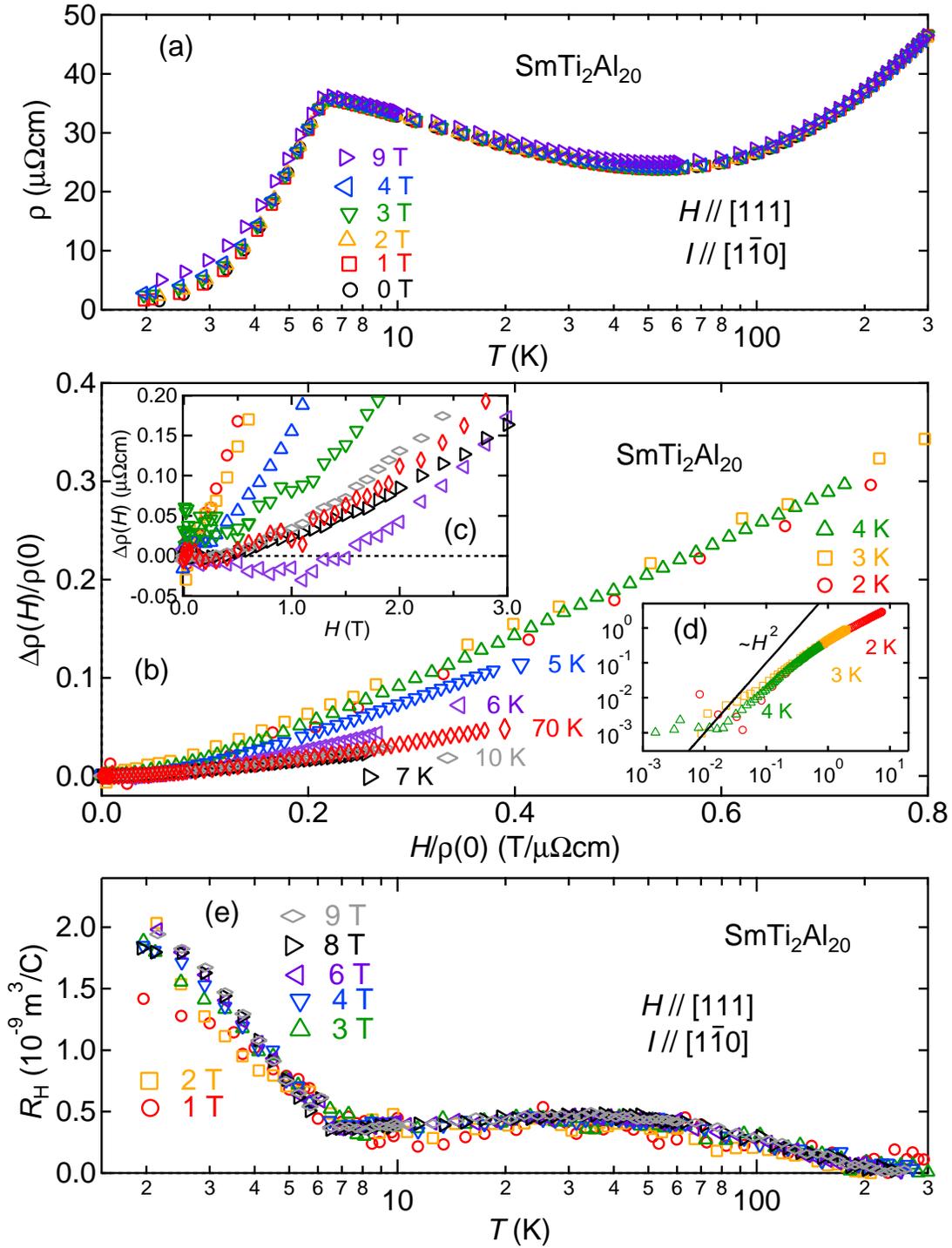}
\end{center}
\caption{(Color online) (a) $T$ dependence of resistivity $\rho$ at various fields using a plate like sample with (111) plane.
The current flows along the [1$\bar{1}$0] direction.
(b) Kohler's plot at various temperatures (2 $\sim$ 70 K).
(c) $H$ dependence of magnetoresistance $\Delta \rho (H)$ at various temperatures (2 $\sim$ 70 K) and below 3 T.
(d) the log-log plot of Kohler's plot below 4 K.
The solid line represents $\sim H^{2}$ line.
(e) $T$ dependence of Hall coefficient $R_{{\rm H}}$ at various fields.
}
\label{f2}
\end{figure}
\quad Fig.2 (a) shows the $T$ dependence of electrical resistivity $\rho$ at various magnetic fields applied along the [111] direction with the current along the [$1\bar{1}0$] direction.
The residual resistivity ratio [RRR = $\rho$(300 K)/$\rho$(2 K)] is about 40, which is larger than those of Yb$T_{2}$Zn$_{20}$ series (3 $\sim$ 30)\cite{07Torika}, indicating the high quality of the present samples.
From RT, $\rho$ decreases monotonically with decreasing temperature and exhibits a minimum around 50 K (= $T_{{\rm min}}$).
Below 50 K, $\rho$ exhibits a remarkable Kondo-like log $T$ dependence down to $T_{x}$ where it shows a sharp decrease.
The resistance increase below $T_{{\rm min}}$, [${\rho(T_{x}) - \rho(T_{{\rm min}})}]/\rho(T_{x})$ = 43 \%, is remarkably larger than that of the typical Sm-based HF system, e.g., SmSn$_{3}$\cite{85Kasaya}.
As was already noticed in $\chi$-$T$ curves (Fig.1 (a)), applied magnetic field has no evident effect on $T_{x}$.
In Fig.2 (a), the effect of magnetic field on $\rho$ is found to be not so simple.
Fig. 2 (b) shows the Kohler's plot, $\Delta \rho (H)/\rho (0) (= [\rho (H) - \rho(0))/\rho(0)]$ versus $H/\rho (0)$, at various temperatures.
The data below 4 K are approximately on a single curve.
However, the downward deviation of the data from the single curve becomes progressively apparent above 5 K.
The downward deviation of $\Delta\rho(H) / \rho (0)$ is ascribable to the negative magnetoresistance (MR) component caused by the suppression of magnetic scatterings of conduction electrons by $H$.
In fact, at 6 K just below $T_{x}$ the negative MR is observed up to 1.5 T in the $H$ dependence of $\Delta \rho(H)$ [Fig.2 (c)].
The enhancement of negative MR at around $T_{x}$ is naturally ascribable to the critical fluctuations of the transition as was commonly observed in various type of phase transitions\cite{Rossiter}.
However, the negative MR remains up to 70 K, which is an order of magnitude higher than $T_{x}$.
This fact indicates that some other contributions (probably Kondo effect) plays a role in the negative MR over the wider temperature range.
\\
\quad In order to obtain information on the Fermi Surface (FS) topology, Kohler's plot of MR below 4 K is made in double-log form as shown in Fig.2 (d).
Since SmTi$_{2}$Al$_{20}$ is a compensated metal with two formula units in a fcc primitive unit cell, the transverse MR is expected to show $H^{2}$ dependence both in the low-field and high-field conditions, unless open orbits exist.
Fawcett reported that the Kohler's plot for various metals can be classified into two types; i.e., $\rho(H)$ for compensated metals shows $H^{2}$-dependence over the investigated field strength, while for uncompensated metals, it initially increases as $H^{2}$ and shows a tendency to saturate around a crossover field $H_{{\rm cr}}$\cite{Faw}.
Above the crossover field, conduction electrons nearly complete their cyclotron motion. 
In the free electron model, $\omega_{c} \tau = H_{{\rm cr}}/ne\rho (0) \sim 1$, where $\omega_{c}$, $\tau$, and $n$ are the cyclotron frequency, the relaxation time and the density of conduction electrons, respectively. 
In the present case, the field dependence of $\Delta \rho/\rho(0)$ below $H/\rho(0) \sim$ 0.1 T/$\mu \Omega$cm is approaching the $H^{2}$ dependence expected for the low field condition.
At higher fields, however, $\Delta \rho(H)/\rho(0)$ remarkably deviates from $H^{2}$-dependence and the exponent is close to 1 at the highest field (at 2 K and 9 T), where $\Delta \rho/\rho(0) \approx 3$ indicates nearly high field condition ($\omega_{c} \tau \ge 1$).
Such field dependence of $\Delta \rho/\rho(0)$ similar to that in Fig.2 (d) has been reported in several compensated metals for $H$ along selected high symmetry axes, such as $H$//[110] in Pb\cite{Pb}.
In those cases, the compensation of carrier numbers between electron-like and hole-like FSs is broken by so-called geometric discompensation\cite{Fick}.
For a singular field direction of high symmetry axis, two aperiodic open orbits possibly intersect and form a closed orbit.
As a result, the effective number of carriers per unit cell changes from that obtained for an equivalent surface under $H$ away from the symmetry axis for which no open orbit exists.
Due to the change in carrier number, $\Delta n$, $\Delta \rho/\rho(0)$ tends to saturate at high fields.
\\
\quad Fig.2 (e) shows the $T$ dependence of Hall coefficient, $R_{{\rm H}} = \rho_{{\rm H}}/H$ under selected magnetic fields.
$R_{{\rm H}} \sim 1 \times 10^{-11}$ m$^{3}$/C at RT is a typical value for compensated metals.
In agreement with the other physical properties, $T_{x}$ is almost independent of magnetic field.
As origins of the change in $R_{{\rm H}}$ below $T_{x}$, both the changes in carrier numbers and in relaxation time are discussed in general.
As a typical example for the former, the disappearance of the main FS due to the superzone gap formation has been reported in P-based filled skutterudites such as PrFe$_{4}$P$_{12}$\cite{PrFe4P12} and PrRu$_{4}$P$_{12}$\cite{PrRu4P12}.
In those cases, both $\rho$ and $R_{{\rm H}}$ exhibits a sharp upturn just below the transition temperature.
In contrast, $\rho$ for SmTi$_{2}$Al$_{20}$ exhibits a steep decrease just below $T_{x}$, ruling out the possibility of a disappearance of the main FSs.
Therefore, the change in scattering mechanism probably plays a main role in the upturn of $R_{{\rm H}}$ in SmTi$_{2}$Al$_{20}$ as reasonably expected from the sharp decrease of magnetic scattering below $T_{x}$.
In magnetic metals, the Hall resistivity is empirically represented by the sum of the normal Hall component $R_{0}H$ and the anomalous Hall component $4\pi R_{s}M$,i.e., $\rho_{{\rm H}} = R_{0}H + 4\pi R_{s}M$.
The results on MR indicate that the anomalous Hall component cannot be negligible in SmTi$_{2}$Al$_{20}$ except at the lowest temperature.
However, it is hard to separate the two components, since the magnetization curve is almost linear at all temperatures investigated.
As a notable feature in Fig.2 (e), the field dependence of $R_{{\rm H}}$ becomes progressively evident with decreasing temperature below $T_{x}$.
At 2 K where the anomalous Hall contribution is quite small as inferred from the results on MR, $\rho_{{\rm H}}$ versus $H$ curve is not linear but has tilted S shape;i.e., $\rho_{{\rm H}}/H$ at the lowest field is smaller than that at 9 T with a peak in the intermediate field (data not shown).
Those behaviors of $R_{{\rm H}}$ at 2 K could be naturally ascribed to the crossover from low-field to high-field condition, consistent with the field dependence of $\Delta\rho/\rho(0)$.
Assuming that $R_{{\rm H}} \sim 2.0 \times 10^{-9}$ m$^{3}$/C  at 2 K under 9 T reflects the discompensated carrier number, the thickness of the discompensated orbit is estimated to be $\sim 0.04 \times 2\pi/a$, where $a$ is the lattice parameter.
In order to confirm this scenario, dHvA experiment along with a band structure calculation is needed.
\\
\begin{figure}
\begin{center}
\includegraphics[width=\linewidth]{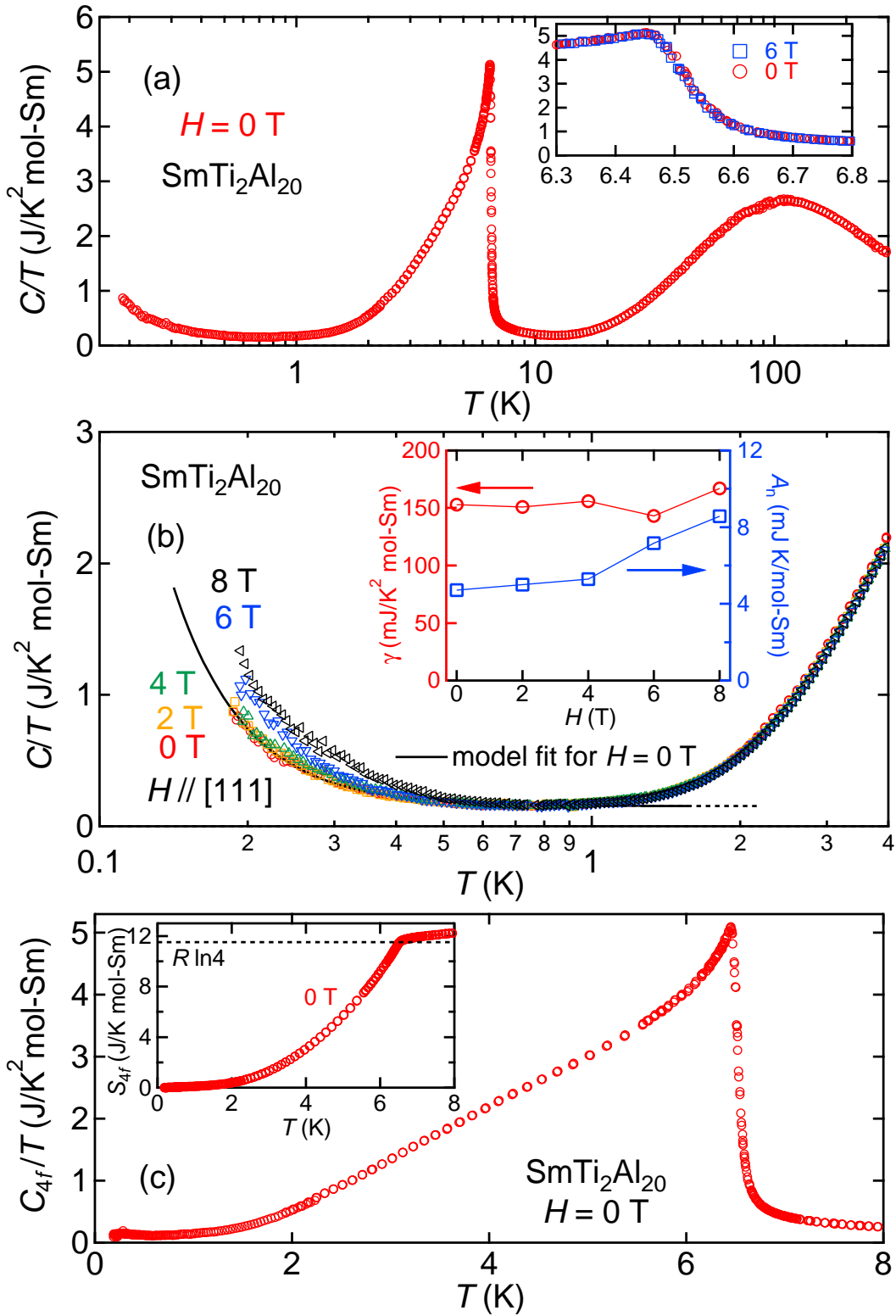}
\end{center}
\caption{(Color online) (a) $T$ dependence of $C/T$ for single crystalline SmTi$_{2}$Al$_{20}$ at 0 T.
Inset: enlarged view at around $T_{x}$ at 0 and 6 T with random crystalline-orientation.
(b) Low temperature part of $C/T$ below 4 K at various fields.
Solid line indicates the fitting line for 0 T below 1 K (see text for details).
Inset: $H$ dependence of $\gamma$ and $A_{{\rm n}}$ estimated from the model fitting.
(c) $T$ dependences of $C_{4f}/T$ and (Inset) $S_{4f}$ at 0 T.
The dashed line represents $R\ln 4$ expected from a $\Gamma_{8}$ quartet ground state.
}
\label{f4}
\end{figure}
\quad In Fig.3 (a), the $T$ dependence of specific heat divided by temperature $C/T$ at 0 and 6 T is shown.
A sharp $\lambda$-type peak appearing at $T_{x}$ provides a thermodynamical evidence for the phase transition.
The $C/T$ data also confirm the field-insensitive nature of $T_{x}$ [the inset of Fig.3 (a)].
Low temperature parts of $C/T$ below 4 K at various magnetic fields are shown in Fig.3 (b).
Below $T_{x}$, $C/T$ steeply decreases and is almost flat around 1 K, where the itinerant quasiparticle excitations dominate. 
An upturn appearing below 0.5 K is derived from the nuclear Schottky contributions $C_{{\rm nu}}$.
As shown in Fig.3 (b), the $C/T$ data below 1 K are fitted by
\begin{equation}
\frac{C}{T} = \gamma + \frac{A_{{\rm n}}}{T^{3}},
\end {equation}
and $\gamma$ = 0.15 J/(K$^{2}$ mol-Sm) and $A_{{\rm n}}$ = 4.7 mJ K/mol-Sm are obtained at 0 T. 
Compared with $\sim$ 0.02 J/(K$^{2}$ mol-Yb) of nonmagnetic reference Yb$T_{2}$Al$_{20}$ \cite{10Higashinaka}, the $\gamma$ value indicates that the quasiparticle mass is enhanced moderately even in the ordered state of SmTi$_{2}$Al$_{20}$ although it is smaller than those for the typical Sm-based HF compounds, i.e., 0.18 J/(K$^{2}$ mol-Sm) for SmSn$_{3}$\cite{87Kasuya}, 0.28 J/(K$^{2}$ mol-Sm) for SmPd$_{3}$\cite{88Liu}, 0.37 J/(K$^{2}$ mol-Sm) for SmFe$_{4}$P$_{12}$\cite{03Takeda} and 0.82 J/(K$^{2}$ mol-Sm) for SmOs$_{4}$Sb$_{12}$\cite{05Sanada}.
The obtained $H$ dependence of $\gamma$ [the inset of Fig.3 (b)] shows extremely weak $H$ dependence, which is similar with that in the field insensitive HF behavior of SmOs$_{4}$Sb$_{12}$\cite{05Sanada}.
\\
\quad For a rough estimation of the 4$f$ electron contribution $C_{4f}$, we tentatively define the phonon contribution $C_{{\rm ph}}$ and the no-$4f$ conduction electron contribution $C_{{\rm el}}$ using those of nonmagnetic YbV$_2$Al$_{20}$ [$\gamma \sim$ 0.02 J/K$^{2}$ mol-Yb], which exhibits Pauli paramagnetic behavior with nonmagnetic Yb$^{2+}$ ions\cite{10Higashinaka}, assuming that the measured specific heat is expressed as $C_{4f} + C_{{\rm el}} + C_{{\rm ph}} + C_{{\rm nu}}$.
Note that $C_{4f}$ includes the contributions of the quasiparticle mass enhancement at low temperatures and CEF excitations at high temperatures when the Kondo temperature $T_{{\rm K}}$ is low enough (see below for confirmation). 
The extracted $C_{4f}/T$ and the associated entropy $S_{4f}$ at 0 T are shown in Fig.3 (c).
$C_{4f}/T$ does not show a simple power law or exponential decrease below $T_{x}$ and there appears a weak shoulder-like anomaly at around 4 K.
In the cubic CEF, the Sm$^{3+}$ $J$ = 5/2 multiplet splits into a $\Gamma_{8}$ quartet and a $\Gamma_{7}$ doublet.
The fact that $S_{4f}$ reaches to $R\ln 4$ at $T_{{\rm x}}$ is consistent with a $\Gamma_{8}$ quartet CEF ground state.
As the -$\log T$ dependence of $\rho(T)$ indicates, Kondo effect is expected to play a role in $S_{4f}(T)$.
Comparing $S_{4f} (T)$ in $T > T_{x}$ with a NCA calculation\cite{07Sakai}, $T_{{\rm K}}$ and the energy separation between $\Gamma_{8}$ and $\Gamma_{7}$ CEF states, $\Delta$, have tentatively been estimated to be about 2 K and 25 K, respectively.
\\
\quad Let us consider the nuclear specific heat constant $A_{{{\rm n}}}$.
The experimental $A_{{\rm n}}$ consists of the contributions from Sm, Ti and Al nuclei ($A_{{\rm n}} \equiv A_{{\rm n}}^{{\rm Sm}} + A_{{\rm n}}^{{\rm Ti}} + A_{{\rm n}}^{{\rm Al}}$).
In zero field, $A_{{\rm n}}^{{\rm Sm}}$ dominates over the others\cite{comment1} since it is largely enhanced by the strong intrasite hyperfine coupling between the nucleus and 4$f$ electrons\cite{63Bleaney}.
The isotopes that contribute to $A_{{\rm n}}^{{\rm Sm}}$ are $^{147}$Sm and $^{149}$Sm: the nuclear spin $I$ = 7/2 (7/2), the natural abundance $n$ = 0.151(0.139) and the magnetic dipole hyperfine coupling constant $A_{{\rm hf}}$ = -0.0116 (-0.0095) K\cite{63Bleaney} for $^{147}$Sm ($^{149}$Sm).
Using the isotope-averaged $\overline{A^{2}_{{\rm hf}}} \equiv \sum_{i} (nA^{2}_{{\rm hf}})_{i}$ = 3.29$\times 10^{-5}$ K$^{2}$, where $i$ runs over the isotopes, $A_{{\rm n}}^{{\rm Sm}}$ is given by
\begin{equation}
A_{{\rm n}}^{{\rm Sm}} = R\overline{A_{{\rm hf}}^{2}} \left( \frac{m_{{\rm Sm}}}{g_{J}} \right)^{2} \frac{I(I+1)}{3},
\end{equation}
where $R$ is the gas constant and $g_{J}$ is the Land$\acute{{\rm e}}$ $g$-factor.
The non zero $A_{{\rm n}}^{{\rm Sm}}$ indicates that non zero Sm dipole moments ($m_{{\rm Sm}} \neq 0$) order below $T_{x}$.
Therefore, $m_{{\rm Sm}}$ can be estimated from $A_{{\rm n}}^{{\rm Sm}}$ as has been done in SmRu$_{4}$P$_{12}$\cite{07Aoki}.
Assuming $A_{{\rm n}} = A_{{\rm n}}^{{\rm Sm}}$, $m_{{\rm Sm}}=0.51 \mu_{{\rm B}}$/Sm is obtained at 0 T\cite{comment2}.
This indicates that the primary order parameter for the transition at $T_{x}$ is the Sm dipole moment.
However, the estimated value of $m_{{\rm Sm}}$ is slightly lower than that of a $\Gamma_{8}$ quartet state ($\sim$ 0.65 $\mu_{{\rm B}}$).
This contraction of the Sm magnetic moment may be caused by the $c$-$f$ hybridization as inferred from the Kondo-like $\log T$ dependence in $\rho(T)$ and the suppressed Curie term in $\chi(T)$.
Another possibility is the octupole degrees of freedom playing a role as a secondary order parameter in the ordering below $T_x$.
The $\Gamma_{8}$ quartet state includes $\Gamma_4$-type octupole moments and they can order parasitically since they have the same symmetry as the dipole moments\cite{97Shiina}.
In such a case, the size of the ordered dipole moment can be smaller than that of the original value as discussed in SmRu$_{4}$P$_{12}$\cite{07Aoki}.
Since the octupole moments do not couple directly with the uniform applied fields, it is expected that the ordered state can be robust against the applied fields to some extent.
This feature might explain the field-insensitive nature of the phase transition at $T_x$ in SmTi$_2$Al$_{20}$.
\\
\quad In summary, we have performed the thermodynamic and transport property measurements of high-quality single crystalline SmTi$_2$Al$_{20}$ with the RRR $\sim$ 40.
We have revealed the field-insensitive phase transition at $T_{x}$ = 6.5 K.
The Kondo-like $\log T$ dependence in $\rho$ and the enhanced $H$-insensitive $\gamma$ [= 0.15 J/(K$^{2}$ mol-Sm)] reflect strong correlations in SmTi$_{2}$Al$_{20}$, indicating that this is a rare example in Sm-based compounds.
The specific heat analysis reveals that the CEF ground state is a $\Gamma_{8}$ quartet state and its Sm dipole moment becomes the primary order parameter in the ordered state ($\sim 0.5 \mu_{{\rm B}}$/Sm at $T \sim 0$).
The $\Gamma_{4}$-type octupole moments included in the $\Gamma_{8}$ quartet state can play a role as a secondary order parameter.
This possibility might explain the field-insensitive nature of the phase transition at $T_x$ in SmTi$_2$Al$_{20}$.
\\
\acknowledgements
This work was supported by a Grant-in-Aid for Young Scientist (B) (No.21740261) and Scientific Research (B) (No.20340094) and (C) (No.23540421) from the Japan Society for the Promotion of Science, and a Grant-in-Aid for Scientific Research on Priority Area ''Ubiquitous'' (No.20045015), and Innovative Areas "Heavy Electrons'' (No.20102007) from MEXT of Japan.
\\


\begin{thebibliography}{9}
\bibitem{02Bauer} E. D. Bauer, N.A. Frederick, P.-C. Ho, V.S. Zapf, and M.B. Maple: Phys. Rev. B {\bf 65} (2002) 100506.
\bibitem{02Matsuhira} K. Matsuhira, Y. Hinatsu, C. Sekine, T. Togashi, H. Maki, I. Shirotani, H. Kitazawa, T. Takamasu and G. Kido: J. Phys. Soc. Jpn. {\bf 71} (2002) Suppl. p. 237.
\bibitem{05Yoshizawa} M. Yoshizawa, Y. Nakanishi, M. Oikawa, C. Sekine, I. Shirotani, S.R. Saha, H. Sugawara and H. Sato: J. Phys. Soc. Jpn. {\bf 74} (2005) 2141.
\bibitem{07Aoki} Y. Aoki, S. Sanada, D. Kikuchi, H. Sugawara and H. Sato: J. Phys. Soc. Jpn. {\bf 76} (2007) 113703.
\bibitem{05Sanada} S. Sanada, Y. Aoki, H. Aoki, A. Tsuchiya, D. Kikuchi, H. Sugawara, and H. Sato: J. Phys. Soc. Jpn. {\bf 74} (2005) 246.
\bibitem{07Torika} M.S. Torikachvili, S. Jia, E.D. Mun, S.T. Hannahs, R.C. Black, W.K. Neils, D. Martien, S.L. Bud'ko and P.C. Canfield: Proc. Natl. Acad. Sci. U.S.A. {\bf 104} (2007) 9960.
\bibitem{08Saiga} Y. Saiga, K. Matsubayashi, T. Fujiwara, M. Kosaka, S. Katano, M. Hedo, T. Matsumoto and Y. Uwatoko: J. Phys. Soc. Jpn. {\bf 77} (2008) 053710.
\bibitem{09Yoshiuchi} S. Yoshiuchi, M. Toda, M. Matsushita, S. Yasui, Y. Hirose, M. Ohya, K. Katayama, F. Honda, K. Sugiyama, M. Hagiwara, K. Kindo, T. Takeuchi, E. Yamamoto, Y. Haga, R. Settai, T. Tanaka, Y. Kubo, and Y. Onuki: J. Phys. Soc. Jpn. {\bf 78} (2009) 123711.
\bibitem{09Jia} S. Jia, Ni Ni, S. L. BudÅfko, and P. C. Canfield; Phys. Rev. B {\bf 80} (2009) 104403.
\bibitem{11Onimaru} T. Onimaru, K.T. Matsumoto, Y.F. Inoue, K. Umeo, T. Sakakibara, Y. Karaki, M. Kubota, and T. Takabatake: Phys. Rev. Lett. {\bf 106} (2011) 177001.
\bibitem{68Krypy} P.I. Krypyakevych and O.S. Zarechnyuk: Dopov. Akad. Nauk Ukr. RSR Ser. A: Fiz. Tekh {\bf 30} (1968) 364.
\bibitem{95Niemann} S. Niemann and W. Jeitschko: J. Solid State Chem. {\bf 114} (1995) 337.
\bibitem{97Nasch} T. Nasch, W. Jeitschko and U. C. Rodewald: Z. Natuurforsch., B: Chem. Sci. {\bf 52} (1997) 1023.
\bibitem{98Thiede} V. M. T. Thiede, W. Jeitschko, S Niemann and T. Ebel: J. Alloys Compd. {\bf 267} (1998) 23.
\bibitem{10Sakai} A. Sakai and S. Nakatsuji: J. Phys. Soc. Jpn. {\bf 80} (2011) 063701.
\bibitem{73Wijn} H.W. Wijn, A.M. van Diepen and K.H.J. Buschow: Phys. Rev. B {\bf 7} (1973) 7.
\bibitem{85Kasaya} M. Kasaya, B. Liu, M. Sera, T. Kasuya, D. Endoh, T. Goto and T. Fujimura: J. Magn. Magn. Mater. {\bf 52} (1985) 289.
\bibitem{Rossiter} P.L. Rossiter: {\it The electrical resistivity of metals and alloys} (Cambridge U. P., 1987).
\bibitem{Faw} E. Fawcett: Adv. Phys. {\bf 13} (1964) 139.
\bibitem{Pb} H. Sato, Y. Imai and K. Yonemitsu: J. Phys. F: Met. Phys. {\bf 13} (1983) 1071.
\bibitem{Fick} F.R. Fickett: Phys. Rev. B {\bf 3} (1971) 1945.
\bibitem{PrFe4P12} H. Sato, D. Kikuchi, K. Tanaka, H. Aoki, K. Kuwahara, Y. Aoki, M. Kohgi, H. Sugawara and K. Iwasa: J. Magn. Magn. Mater. {\bf 310} (2007) 188.
\bibitem{PrRu4P12} S.R. Saha, H. Sugawara, T. Namiki, Y. Aoki, and H. Sato: Phys. Rev. B {\bf 80} (2009) 014433.
\bibitem{10Higashinaka} R. Higashinaka, A. Nakama, M. Ando, M. Watanabe, Y. Aoki and H. Sato: J. Phys.: Conf. Ser. {\bf 273} (2011) 012033.
\bibitem{87Kasuya} T. Kasuya, M. Kasaya, K. Takegahara, F. Iga, B. Liu and N. Kobayashi: J. Less Comm. Metals {\bf 127} (1987) 337.
\bibitem{88Liu} B. Liu,  M. Kasaya and T. Kasuya: J. de Physique Colloque {\bf C8}, Supplement (1988) 369.
\bibitem{03Takeda} N. Takeda and M. Ishikawa: J. Phys.: Condens. Matter {\bf 15} (2003) L229.
\bibitem{07Sakai} For a NCA calculation, we used the CXcal-excel program developed by O. Sakai and H. Kitazawa.
\bibitem{comment1} First, the contributions from Ti nuclei can be neglected since $A_{{\rm n}}^{{\rm Ti}}$ is three orders of magnitude smaller than $A_{{\rm n}}^{{\rm Al}}$. Second, the observed $A_{{\rm n}}$ cannot be accounted for by $A_{{\rm n}}^{{\rm Al}}$ since it requires the local field at Al nuclear site to be over 7 T, which is unrealistic because, even if the Sm magnetic moment is fully polarized ($\sim 0.8$ $\mu_{{\rm B}}$), the induced dipole field at Al nuclear site is only $\sim0.05$ T.
Therefore, we assume $A_{{\rm n}} \simeq A_{{\rm n}}^{{\rm Sm}}$ at 0 T.
\bibitem{63Bleaney} B. Bleaney: J. Appl. Phys. {\bf 34} (1963) 1024.
\bibitem{comment2} It is difficult to obtain the $H$ dependence of $m_{{\rm Sm}}$ because $A_{{\rm n}}^{{\rm Al}}$ increases with increasing $H$ and becomes the dominant factor in $A_{{\rm n}}$ above $\sim$ 3 T.
For further investigation for the $H$ dependence of $m_{{\rm Sm}}$, NMR study on the Al sites is essential.
\bibitem{97Shiina} R. Shiina, H. Shiba and P. Thalmeier: J. Phys. Soc. Jpn. {\bf 66} (1997) 1741.
\end{thebibliography}
\end{document}